\documentclass[prd,onecolumn,preprintnumbers,nofootinbib]{revtex4}

\usepackage{graphicx}

\usepackage[plainpages=false, colorlinks=true, anchorcolor=blue, linkcolor=blue, citecolor=blue, bookmarks=false]{hyperref}
\usepackage{color}
\usepackage{float}
\usepackage{amsfonts,amsmath,amssymb}
\usepackage{natbib}
\usepackage{array}
\usepackage{subcaption}
\usepackage{caption}
\newcommand{\rthis}[1]{\textcolor{black}{#1}}
\begin{document}
\newcolumntype{P}[1]{>{\centering\arraybackslash}p{#1}}
\pdfoutput=1
\newcommand{\jcap}{JCAP}
\newcommand{\araa}{Annual Review of Astron. and Astrophys.}
\newcommand{\apss}{Astrophysics and Space Sciences}
\newcommand{\aj}{Astron. J. }
\newcommand{\mnras}{MNRAS}
\newcommand{\apjl}{Astrophys. J. Lett.}
\newcommand{\apjs}{Astrophys. J. Suppl. Ser.}
\newcommand{\aap}{Astron. \& Astrophys.}
\renewcommand{\arraystretch}{2.5}
\title{A stacked search for spatial coincidences  between IceCube neutrinos and radio pulsars}
\author{Vibhavasu \surname{Pasumarti}}
\altaffiliation{E-mail:ep20btech11015@iith.ac.in}

\author{Shantanu \surname{Desai}}
\altaffiliation{E-mail: shntn05@gmail.com}

\begin{abstract}
We carry out a stacked search for  spatial coincidences   between all the known radio  pulsars and TeV neutrinos from the IceCube 10 year (2008-2018)  muon track data,  as a followup to our previous work on searching  for  spatial coincidences with individual pulsars.
We consider three different weighting schemes to stack the contributions from each pulsar.  We do not find a statistically significant excess using this method. We report the  95\% c.l.  neutrino flux upper  limit as a function of  the  neutrino energy. We have also made our analysis codes publicly available.

\end{abstract}

\affiliation{Dept  of Physics, IIT Hyderabad,  Kandi, Telangana-502284, India}

\maketitle
\section{Introduction}
The origin of majority of the diffuse neutrino flux in the TeV-PeV energy  range detected by IceCube in 2013~\cite{IceCubescience} is still  unknown~\cite{Halzen23}.
Understanding this origin could also help unravel the source of ultra-high energy cosmic rays. 
Although, searches by the IceCube collaboration have shown evidence for neutrino emission from some  point sources such as  NGC 1068, TXS 0506+056, the majority of IceCube events cannot be attributed to any astrophysical sources~\cite{IceCubedata}. 
 A number of extragalactic  sources have been considered which could contribute to the diffuse neutrino flux,  such as  blazars and other types of AGNs, star-forming
galaxies, FRBs, GRBs, galaxy clusters, other ancillary extragalactic sources in Fermi-LAT catalog, etc by both the IceCube collaboration, and others using the publicly available IceCube dataset. A non-exhaustive list  includes searches for correlation with extra-galactic sources such as  AGNs and GRBs~\cite{Kamionkowski,Hooper,LuoZhang,Smith21,Li22,Icecube10,IceCubeGRB,IceCubeAGN,IceCubeblazars}, FRBs~\cite{Zhang21,Desai23,IceCubeFRB}, high energetic events from the Fermi-LAT catalog~\cite{Li22,IceCubeAGN2}, and  galaxy catalogs using the 2MASS survey~\cite{IceCube2MASS}.

Most recently,  the IceCube collaboration has found a 4.5$\sigma$ evidence for neutrino emission from the Galactic plane using the cascade events~\cite{Science}. Although the observed signal is consistent with diffuse emission from the  galactic plane, some of the signal could arise from a population of unresolved point sources such as supernova remnants, pulsar wind nebulae or unidentified TeV Galactic sources~\cite{Science}.

In a similar vein, a galactic contribution to the all-sky  IceCube diffuse neutrino flux cannot be excluded and has been estimated to be about 10-20\%~\cite{Palladino}. 
Therefore, searches for coincidences with multiple galactic sources have also been carried out~\cite{Lunardini}.  One such source includes pulsars, which are rapidly rotating neutron stars which emit pulsed radio emission~\cite{Reddy}.  Pulsars have large magnetic ($\sim 10^{14}$ G) and induced electric fields,  and because of their large rotational kinetic energy have long been considered as promising sources of TeV (and higher energy) neutrinos,  which could be detected in IceCube and other $\rm{km^2}$ detectors~\cite{Helfand,Burgio,Fang,Fang16,Dey21}. We briefly recap some of the literature on this. The very first calculation of neutrino emission from pulsars, showed that detectors such as DUMAND (proposed in the 1970s), would detect neutrino emission from 1-2 pulsars with neutrino luminosity of $(60-10000) \times 10^{30}$ ergs/sec~\cite{Helfand}. Then, a model for neutrino emission from  young neutron stars with age $\leq 10^5$ years was proposed,   where protons are accelerated in the neutron star magnetosphere  to sufficiently high energies, so that they reach the photo-meson production resonance, which will produce neutrinos with energies of around 50 TeV~\cite{Burgio}. The event rate in IceCube for nine  neutron stars (most of which are also detected as radio pulsars) was estimated to be between $ 10-100 \rm{~km^{-2}~yr^{-1}}$~\cite{Burgio2}. Most intriguingly, a combined analysis of Super-Kamiokande and MACRO dataset showed that one of these sources, viz. PSR B1509-58 showed a statistically significant excess of spatial coincidence corresponding to  2.6$\sigma$ significance~\cite{Desai22}. 
Nevertheless, we could not confirm this excess with the IceCube 10-year muon track data~\cite{Pasumarti_2022}. The  total diffuse neutrino flux of all pulsars using the mechanism proposed in ~\cite{Burgio} was estimated to be $\sim 10^{-7} \rm{GeV cm^{-2} s^{1} sr^{-1}}$~\cite{Jiang07}. However, it was pointed out that if one takes into account the polar cap geometry in the model proposed in ~\cite{Burgio}, the event rates are much more pessimistic~\cite{Bhadra}.
Models for neutrino emission at PeV  energies from young millisecond pulsars from lepton-lepton and lepton-hadron interactions have also been proposed~\cite{Dey21}.  Similarly, there are also models for neutrino emission from fast-spinning newly born extra-galactic  pulsars  in the 0.1-1 EeV energy range~\cite{Fang14}. Nevertheless, the pulsar emission mechanism is still not completely understood~\cite{Melrose}, and there could be other mechanisms of neutrino production hitherto undiscovered.
Hence, it is imperative to carry out a search for neutrino emission from pulsars, which motivated our previous search~\cite{Pasumarti_2022} for spatial coincidences between IceCube neutrinos and radio pulsars located in our galaxy using the 10-year publicly available IceCube muon track data~\cite{IceCubedata}.
However, we could not detect any statistically significant neutrino emission from PSR B1509-58  or any other pulsar~\cite{Pasumarti_2022}. Nevertheless, a stacked search from the combined pulsar population would  enhance any putative signal to noise ratio, compared to a single source.

Therefore, in this work, as  a continuation to our previous work~\cite{Pasumarti_2022}, we perform a stacked search to determine the spatial correlation between IceCube neutrinos from the 10-year muon-track data~\cite{IceCubedata} and radio pulsars in the ATNF catalogue~\cite{ATNF}.  This manuscript is structured as follows. The neutrino and the radio pulsar dataset  are described in Section~\ref{sec:dataset}. The analysis and results are discussed in Section~\ref{sec:analysis} and Section~\ref{sec:results}, respectively. We conclude in Section~\ref{sec:conclusions}.

\section{Dataset}
\label{sec:dataset}
\begin{figure}[h]
    \centering
    \includegraphics[width=\linewidth]{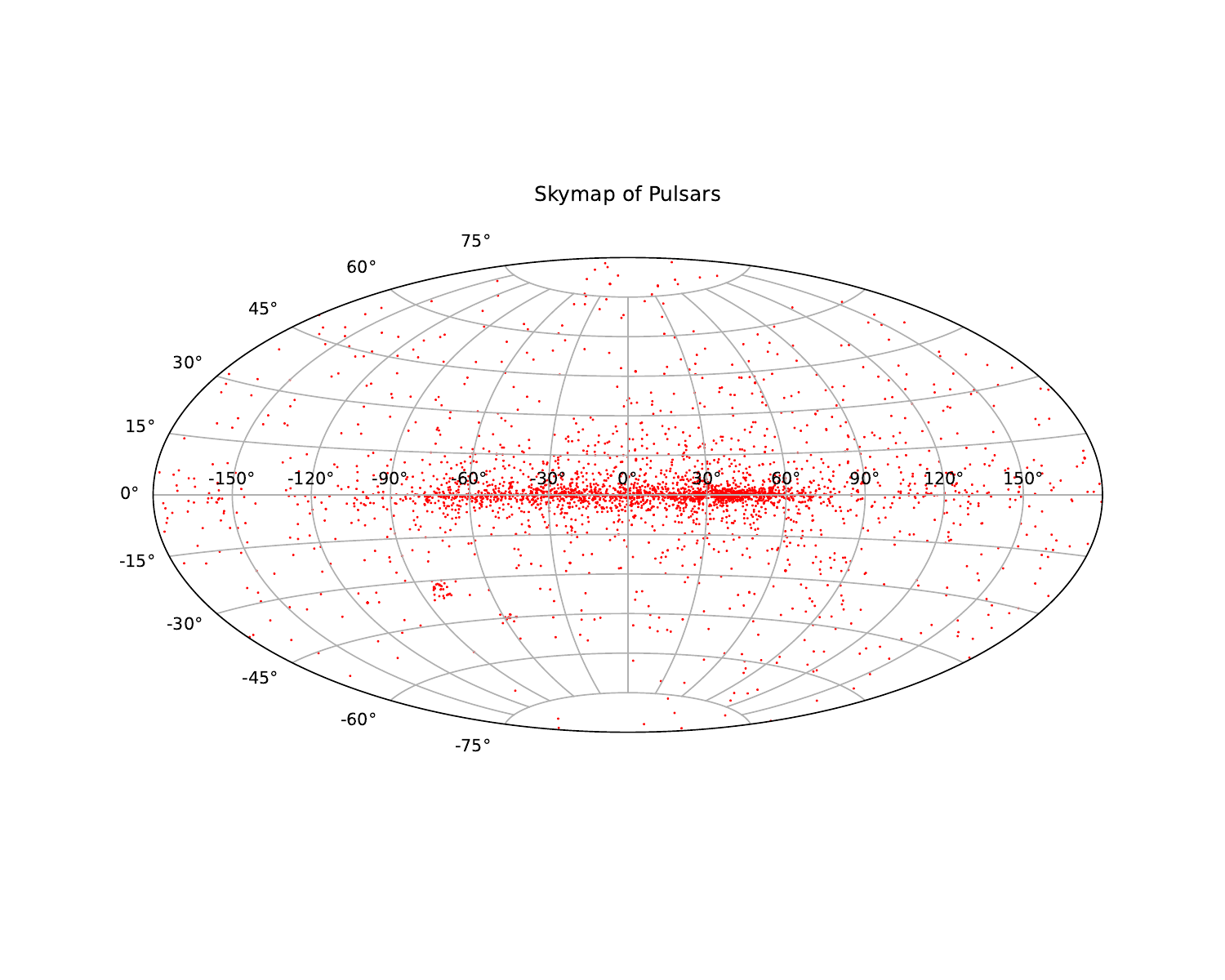}
    \caption{Distribution of  Pulsars in galactic coordinates using Aitoff projection}
    \label{fig:skymap}
\end{figure}

IceCube is a neutrino detector located at the South Pole. It detects neutrinos through the Cherenkov light emitted by the  leptons created by the charged-current interaction of the neutrinos with the surrounding ice in the detector as well as through neutral current interactions.
The IceCube 10-year muon-track data~\cite{IceCubedata} consists of  1,134,550  neutrino events between April 2008 - July 2018  from four different phases of the experiment, each having different livetime. The public release data also contains the detector sensitivity information  as a function of  energy and location on the sky.  The data provided for  every neutrino event consists of  RA, Declination ($\delta$), angular position error, and the reconstructed  muon energy. However, an independent search for dimuons in this  10 year catalog found 38 double-counted events, due  to an error in the reconstruction,  because of which some single muons were masquereded as  two separate muons~\cite{Beacom}. For our analysis,  we used an updated catalog which removes these double-counted events.\footnote{ This dataset with duplicates removed was kindly provided to us by Dr. Bei Zhou.}
The ATNF pulsar catalogue (v1.70)~\cite{ATNF}  consists of 3389 pulsars, whose sky distribution can be found in Fig.~\ref{fig:skymap}.
For our  analysis  we only choose pulsars with galactic latitude  $\pm 15^{\circ}$. For some of our analysis, we also need to know the pulsar  distance and flux at 1400 MHz ($S_{1400}$). 
Furthermore, we also  skip pulsars with $\delta> 85^{\circ}$, since it is difficult to get a robust background estimate for these pulsars. Similar culling of sources near $\delta=90^{\circ}$ has also been done in Ref.~\cite{Kamionkowski}. 

\section{Stacking Analysis}

To implement the  stacking analysis, we extend the unbinned maximum likelihood method by combining   the signal PDFs of  all the sources into a single stack. This way the test statistic becomes more sensitive, since any minor associations  from individual pulsars could  stack up to enhance the significance. For this purpose we follow the same methodology as ~\cite{Kamionkowski,Li22,LuoZhang,Hooper} (based on the implementation first proposed in Ref.~\cite{Montaruli}), 
which have also done a stacking search between neutrinos from  the public IceCube catalog and various sources such as AGNs, FRBs, Fermi-LAT point sources. 

\begin{figure}[H]
    \begin{minipage}{.48\textwidth}
        \centering
        \includegraphics[width=\linewidth]{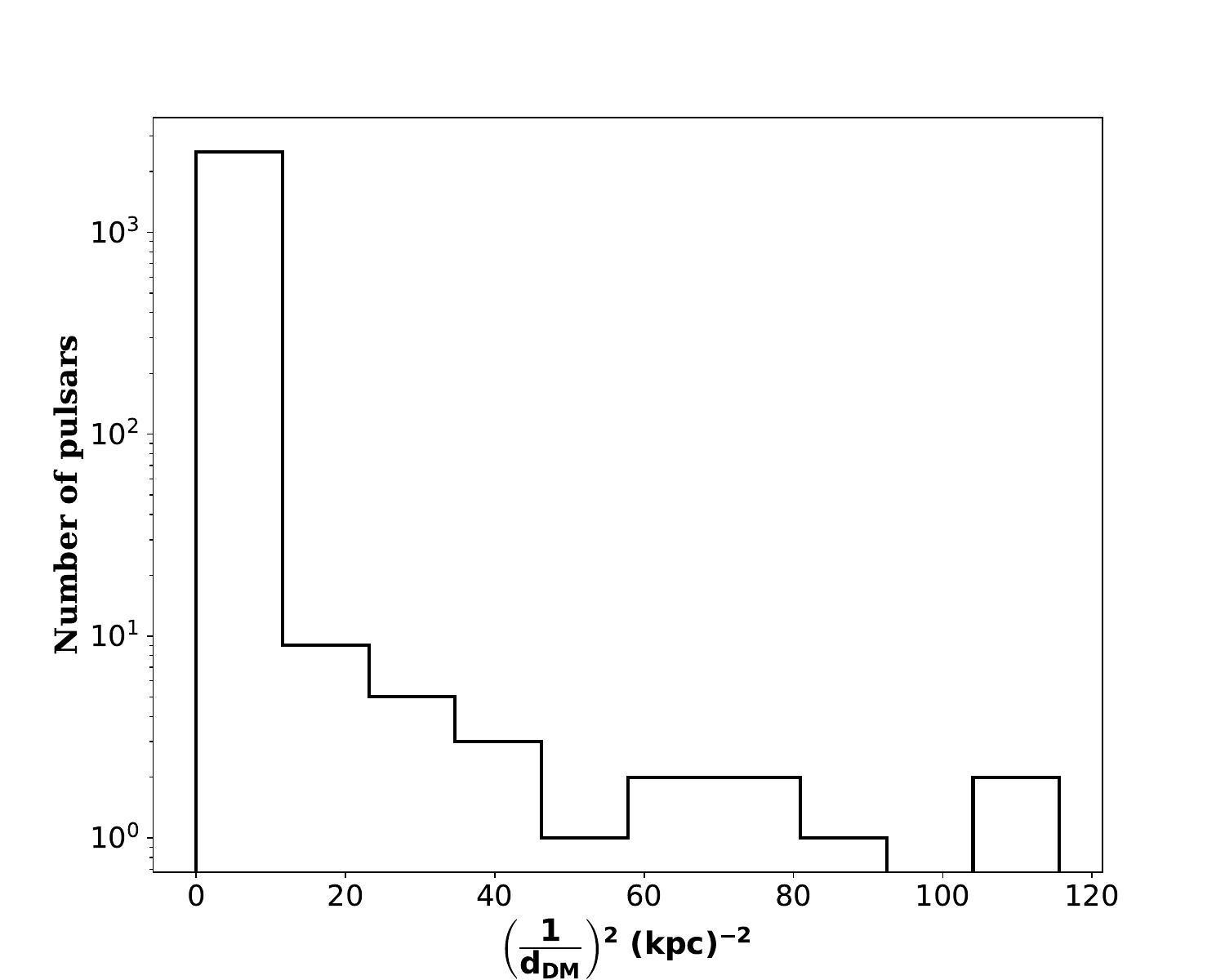}
        \caption{Histogram of $w_{model} = \frac{1}{d_{DM}^2}$ for pulsars with galactic latitude within $\pm 15^{\circ}$, where $d_{DM}$ is the distance estimated from the dispersion measure.}
        \label{fig:ddm2hist}
    \end{minipage}\hspace{1em}
    \begin{minipage}{.48\textwidth}
        \centering
        \includegraphics[width=\linewidth]{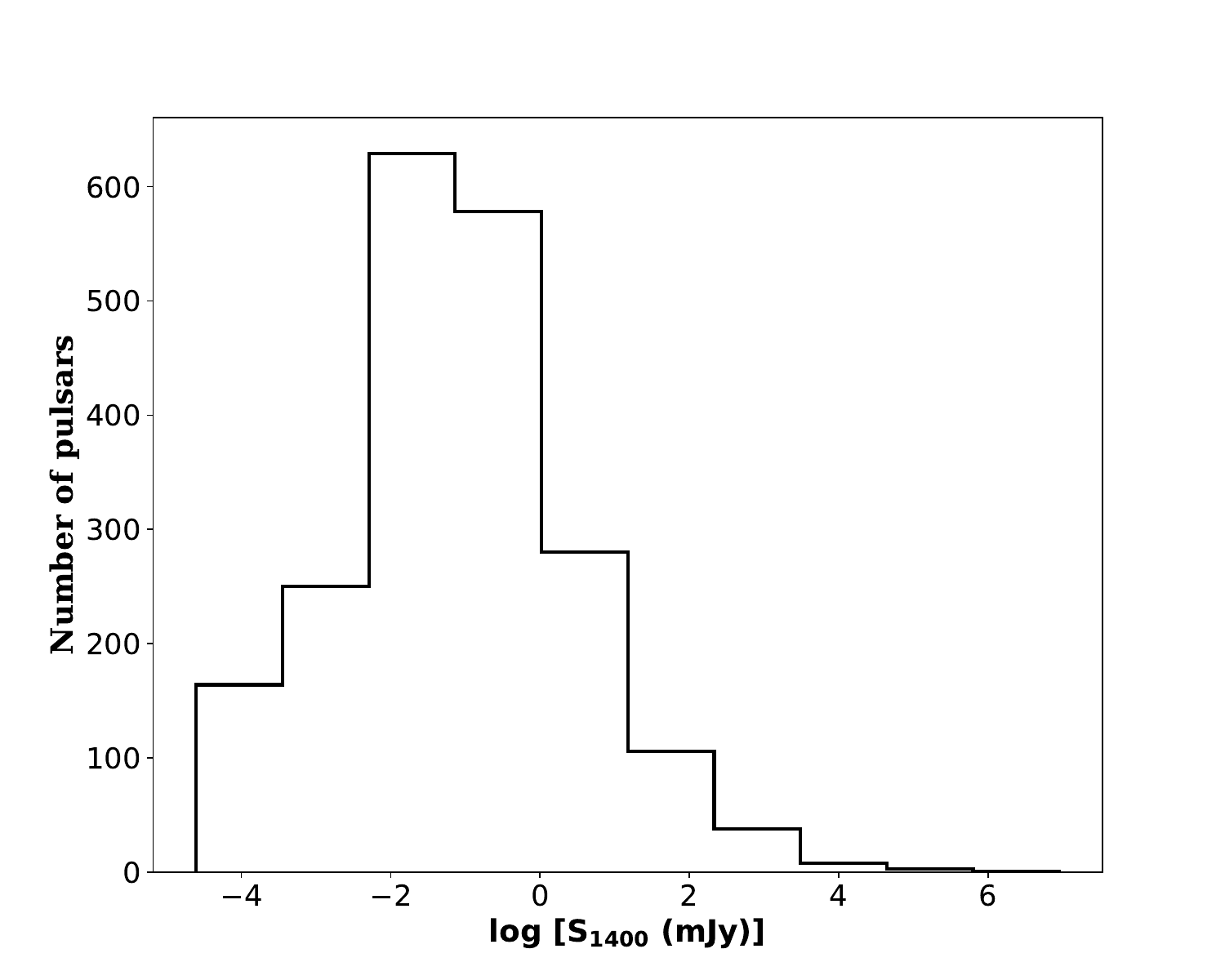}
        \caption{Histogram of $w_{model}=S_{1400}$ for pulsars with galactic latitude within $\pm 15^{\circ}$. The one outlier on the extreme right is for the pulsar  B0833-45, for which  $S_{1400} = 1050$  mJy, compared to the median value of 0.34 mJy.}
        \label{fig:s1400hist}
    \end{minipage}
\end{figure}

If $n_s$ signal events are associated with a pulsar in a dataset of $N$ events, the probability density of an individual event $i$ is given by:
\begin{equation}
	P_i = \frac{n_s}{N} S_i + (1-\frac{n_s}{N}) B_i , 
	\label{eq:prob}
\end{equation}
where $S_i$ and $B_i$ represent the signal and background PDFs, respectively. The likelihood function $\mathcal{L}(n_s)$ for the entire dataset can be expressed as the product of the individual likelihood functions:
\begin{equation}
    \mathcal{L}(n_s) = \prod_i P_i, 
\end{equation}
where the index $i$ extends over all neutrinos in the dataset.
The background PDF ($B_i$) is determined by the solid angle within  $5^{\circ}$ declination band   around  each event $i$ ($\Omega_{\delta_i \pm 5^{\circ}}$):
\begin{equation}
    B_i=\frac{\mathcal{N}_i}{N\Omega_{\delta_i \pm 5^{\circ}}}, 
    \label{eq:Bi}
\end{equation}
where $\mathcal{N}_i$ is the number of events within the $\pm 5^\circ$ declination band of event $i$. In Appendix A, we also redo the analysis  with different declination bands to see if it makes a difference.
In the scenario where there are multiple sources, the signal PDF $S_i$ is a weighted average of the signal PDFs ($S_{ij}$) of all the sources:
\begin{equation}
    S_i = \dfrac{\sum_j \omega_{acc, j}\omega_{model, j} S_{ij}}{\sum_j \omega_{acc, j}\omega_{model, j}}, 
\end{equation}
\begin{equation}
    \text{where } \quad S_{ij} = \frac{1}{2\pi\sigma_i^2}e^{-(|\theta_i-\theta_j|)^2/2\sigma_i^2}. 
    \label{eq:Sij}
\end{equation}
In Eq.~\ref{eq:Sij}, $|\theta_i-\theta_j|$ is the angular distance between the  pulsar and the neutrino,  $\sigma_i$ is the angular uncertainty in the neutrino position, expressed in radians,
$\omega_{acc, j}$ denotes  the weights corresponding to the detector acceptance for the source $j$. The index $j$ is summed over all pulsars in the catalog. 
For a given spectral index $\Gamma$, $\omega_{acc, j}$ is given by:
\begin{equation}
    \omega_{acc,j} = T \times \int A_{eff}(E_{\nu}, \delta_j)E_{\nu}^{\Gamma} dE_{\nu} ,
    \label{eq:weights}
\end{equation}
where $\delta_j$ is the declination of the source, $T$ is the total detector uptime. 
In Eq.~\ref{eq:weights}, $\omega_{model, j}$ corresponds to the signal weight used.
We note that there is no guidance from theoretical models regarding how the neutrino flux scales with the pulsar observables. We therefore choose three different weighting methods, which  have previously been used  in searches for neutrinos from AGNs~\cite{Kamionkowski,Li22}, which are enumerated below:
\begin{enumerate}
    \item $\omega_{model, j}$ = 1: Uniform Weighting. 
    \item $\omega_{model, j} \propto \dfrac{1}{d_{DM}^2}$ where $d_{DM}$ is the distance to the pulsar based on the YMW16 electron density model~\cite{YMW16}. The distribution of  $\dfrac{1}{d_{DM}^2}$ for all the pulsars used in our analysis can be found in Figure~\ref{fig:ddm2hist}.
    \item $\omega_{model, j} \propto S_{1400}$ where $S_{1400}$ is the mean pulsar flux density at 1400 MHz. $S_{1400}$ is available for most pulsars and is the most widely used proxy for the pulsar flux~\cite{Bagchi13}. The distribution of  $S_{1400}$ for all pulsars used for our analysis  can be found in Figure~\ref{fig:s1400hist}. There is one outlier in the distribution of $S_{1400}$ due to the   pulsar (viz. B0833-45), for which  $S_{1400}=1050$ mJy, which is about 3000 times larger than the median $S_{1400}$ value of 0.34 mJy.
\end{enumerate}

We note that for $w_{model}=1$, we used all the 2627 pulsars with galactic latitude within $\pm 15^\circ$, since we do not require ancillary information from the ATNF catalog. However, for the latter two cases, we used 2060 pulsars for which both $S_{1400}$ and distance estimates were available. Note that  there are a total of four pulsars which have $S_{1400}$ estimates,  but no distance estimates. Although we only need $S_{1400}$ for the third weighting scheme, we have skipped these  four pulsars and used the same pulsars as those used for the second weighting scheme.




The expected number of  signal events coming from the source  (${n}_s$)  is given by:
\begin{equation}
    {n}_s = \dfrac{\sum_j \omega_{model, j} \hat{n}_{sj} }{\sum_j \omega_{model, j}}
    \label{eq:ns}
\end{equation}
where
\begin{equation}
    \hat{n}_{s_j} =  T \times \int A_{eff}(E_{\nu}, \delta_j)\frac{dF}{dE_{\nu}}dE_{\nu},
    \label{eq:ns_HAT-no_source}
\end{equation}
where $\hat{n}_{sj}$ is the number of signal events coming from the source indexed by $j$, and $\dfrac{dF}{dE_{\nu}}$ is the expected neutrino spectrum from the source, which  can be modelled using a power-law as follows:
\begin{equation}
    \dfrac{dF}{dE_{\nu}} = \phi_0  \left( \dfrac{E_{\nu}}{100 \text{ TeV}}\right)^{\Gamma}
    \label{eq:flux}
\end{equation} 
where  $\phi_0$ is the flux normalization and $\Gamma$ is the spectral index.
We use the following spectral indices: $\Gamma$ = -2.0, -2.53, -3 (similar to Refs.~\cite{Kamionkowski,Li22})  to calculate the model predicted number of events. Note however that this choice is  arbitrary, and it is straightforward to generalize our results to any other spectral index. The test statistic ($TS$) can be written as follows~\cite{Kamionkowski,Li22,LuoZhang}:
\begin{equation}
    \text{TS} = 2\log\left[\dfrac{\mathcal{L}({n}_s)}{\mathcal{L}(0)}\right], 
\label{eq:TS}    
\end{equation}
\label{sec:analysis}
\noindent where the denominator corresponds to the background or null hypothesis of no signal.  
If the null hypothesis is true,  the distribution of $TS (n_s)$ is given by the   $\chi^2$ distribution for one degree of freedom, according to Wilks' theorem~\cite{Wilks}.  However, for single-source searches, where $n_s$ is obtained by maximization of Eq.~\ref{eq:TS}, there could also be additional excess due to the superposition of a $\delta$-function and Gaussian distribution in case $n_s$ is close to the physical boundary~\cite{Wolf}.
The detection significance  can be quantified by  $\sqrt{TS}$.  In Appendix B, we   demonstrate  that this procedure can recover a true signal with a very high statistical significance.


\section{Results}
\label{sec:results}
\begin{figure}[H]
    \centering
    \includegraphics[width=\textwidth]{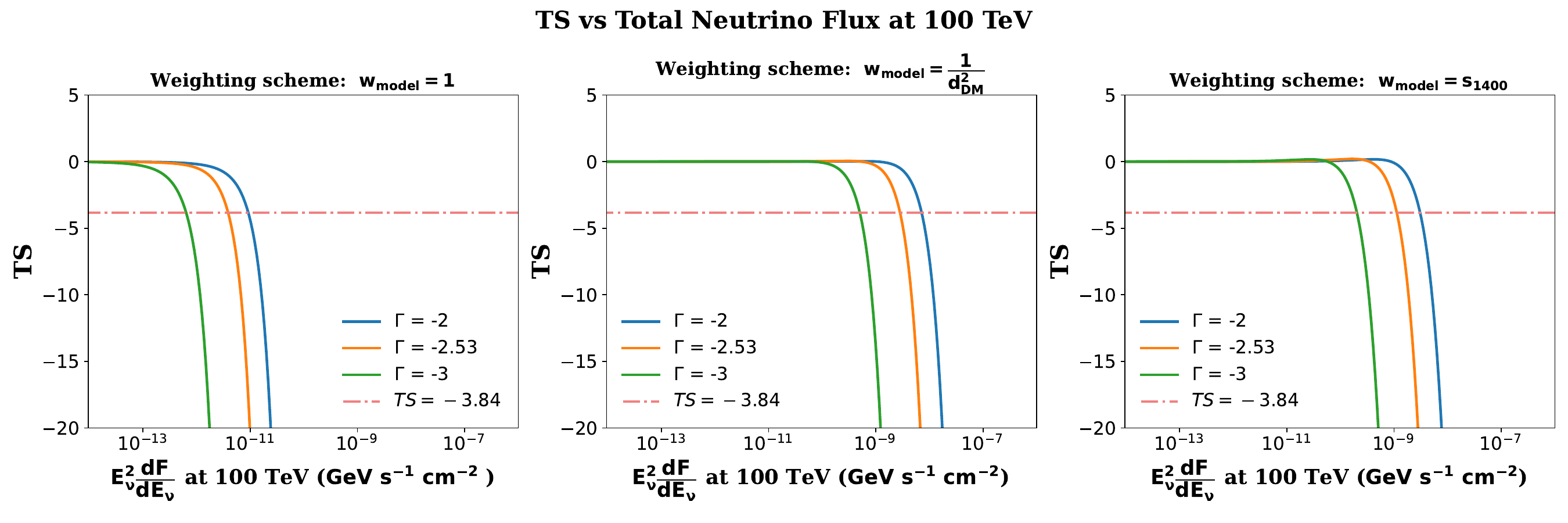}
    \caption{Plot of $TS$ (defined in Eq.~\eqref{eq:TS}) as a function of the total neutrino flux for pulsars in the ATNF catalog,  using the  IceCube 10-year muon-track data. The left panel corresponds to the analysis, where each pulsar is given  the same weight ($w_{model}$). For the middle panel, each pulsar is  weighted  inversely according to the square of its distance, and the rightmost panel corresponds to the case when each pulsar is  scaled based on its flux density at 1400 MHz. For each of the aforementioned analysis, we have considered three different neutrino spectral indices ($\Gamma$).}
    \label{fig:TS_spl}
\end{figure}

We now present the results of our analyses.
We plot $TS$ as a function of the total neutrino flux at 100 TeV  for the three weighting schemes (discussed in Sect.~\ref{sec:analysis})  in Figure~\ref{fig:TS_spl}. 
The graph of $TS$ as a function of the total neutrino flux reveals no statistically significant signal from the sources. Therefore, we conclude that there is no statistically significant spatial excess,  when we do a spatial search  by stacking all  pulsars  in the ATNF catalog with galactic latitude within $|l|<15^{\circ}$.

We then  proceed to calculate  the upper limits on the total contribution by the pulsars at 95\% c.l. At each energy, we calculate the total differential neutrino energy  flux ($E^2 \frac{dF}{dE}$) and plot their 95\% c.l. upper limits for which $TS = -3.84~$\cite{Kamionkowski}. We multiply the upper limits by a factor of three~\cite{Li22,Kamionkowski} since we are  considering the flux due to  the three different flavors of neutrinos. 
For the spectral index $\Gamma = -2.53$, the 95\% c.l.  upper limits on the differential neutrino energy flux ($E_{\nu}^2 \frac{dF}{dE_{\nu}}$) at 100 TeV are $1.2 \times 10^{-11}$ ($\omega_{model} = 1$), $8.5 \times 10^{-9} (\omega_{model} \propto \dfrac{1}{d_{DM}^2})$ and $3.4 \times 10^{-9}$ ($\omega_{model} \propto S_{1400}$) $\rm{GeV~s^{-1}~cm^{-2}}$.
We note that these upper limits correspond to  cumulative limits obtained by summing  over all the  sources.
 Therefore, the upper limits vary considerably between uniform weights and the other two  weighting schemes, because of the large number of pulsars when using uniform weights. The total contribution of pulsars to the diffuse neutrino flux measured using 9.5 years of IceCube data~\cite{IceCubeallsky} is at most 0.003\%, 2.3\%, and 0.9\% respectively for the three  aforementioned weights used.
 The corresponding contribution to the diffuse neutrino flux from the Galactic plane measured by IceCube~\cite{Science} is at most 0.02\%, 13\%, and 5.2\%  for the three aforementioned  weights.

\begin{figure}[H]
    \centering
    \includegraphics[width=\textwidth]{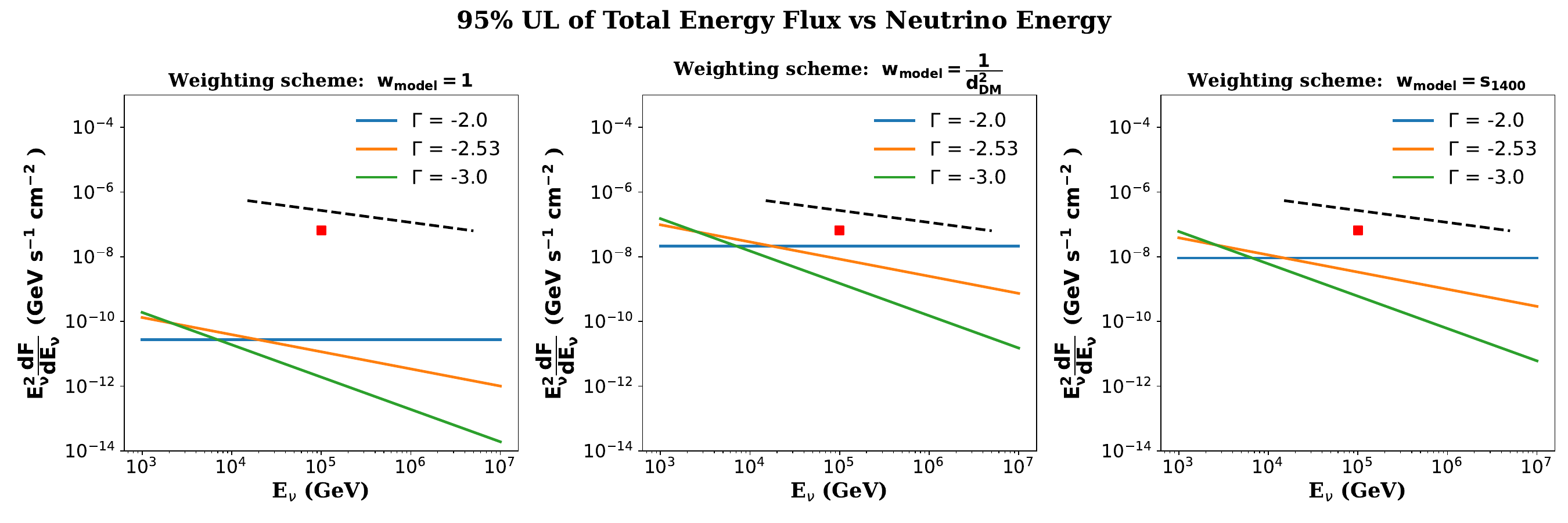}
    \caption{\label{fig:UL_spl} 95\% c.l. upper limit on differential neutrino energy flux as a function of neutrino energy for three different values of $w_{model}$ and for three different spectral indices. The red square corresponds to  the  observed diffuse  neutrino flux from the Galactic plane~\cite{Science} and the black dashed line corresponds to the all-sky diffuse flux obtained by IceCube using 9.5 years of data~\cite{IceCubeallsky}. Note that the IceCube limits~\cite{Science,IceCubeallsky} have been multiplied  by a factor of three to account for all the three flavors of neutrinos. \rthis{We note that for the $w_{model}=1$ scheme, the upper limits are much lower than the other two due to the larger number of pulsars used in the analysis for this weighting scheme. The IceCube  limits from the galactic plane have only been shown at 100 TeV, since only the best-fit flux normalization at 100 TeV has been reported in ~\cite{Science}.}}
\end{figure}
These upper limits on different neutrino energy flux as a function of neutrino energy for the three signal weighting schemes and three spectral indices (for each weighting scheme)
can be found in Figure~\ref{fig:UL_spl}.

\section{Conclusions}
\label{sec:conclusions}
In a previous work~\cite{Pasumarti_2022}, we have performed a single source unbinned likelihood analysis to determine the spatial correlation between IceCube neutrinos and ATNF catalog of pulsars.
In this work, we extend that analysis by doing a stacking analysis~\cite{Kamionkowski,Li22}, in which the signal PDFs of all the sources are combined into a single stack. We then calculate $TS$ (cf. Eq.~\ref{eq:TS}) as a function of the  total neutrino flux at 100 TeV for three different signal weighting schemes and three spectral indices for each weighting method. These plots for all the three signal weights can be found in Figure~\ref{fig:TS_spl}. We find no statistically significant excess from the pulsars in our analysis. We repeat the analysis with different declination bands and find no excess originating due to the pulsars. Therefore, we conclude that there is no statistically significant correlation between the TeV neutrinos detected by IceCube and the stacked contribution from radio pulsars. We then calculate  the 95\% c.l. upper limits on the total neutrino flux of the pulsars, which can be found  in Figure~\ref{fig:UL_spl}.


In the spirit of open science, we have made our analysis codes and the data used for our analysis  publicly available. This  can be found at \url{https://github.com/DarkWake9/IceCube-Package}

\section*{Acknowledgements}
We are grateful to Rong-Lan Li, Yun-Feng Liang,  and Bei Zhou for useful correspondence about their works and also to Bei Zhou for pointing to us about the duplicate events in the IceCube muon track data.   We also thank the anonymous referee for several useful comments and feedback on our manuscript. We acknowledge National Supercomputing Mission (NSM) for providing computing resources of ‘PARAM SEVA’ at IIT, Hyderabad, which is implemented by C-DAC and supported by the Ministry of Electronics and Information Technology (MeitY) and Department of Science and Technology (DST), Government of India. 
\bibliography{main}

\section*{Appendix A: Tests with different declination bands}
\begin{figure}[H]
    \centering
    \includegraphics[width=\textwidth]{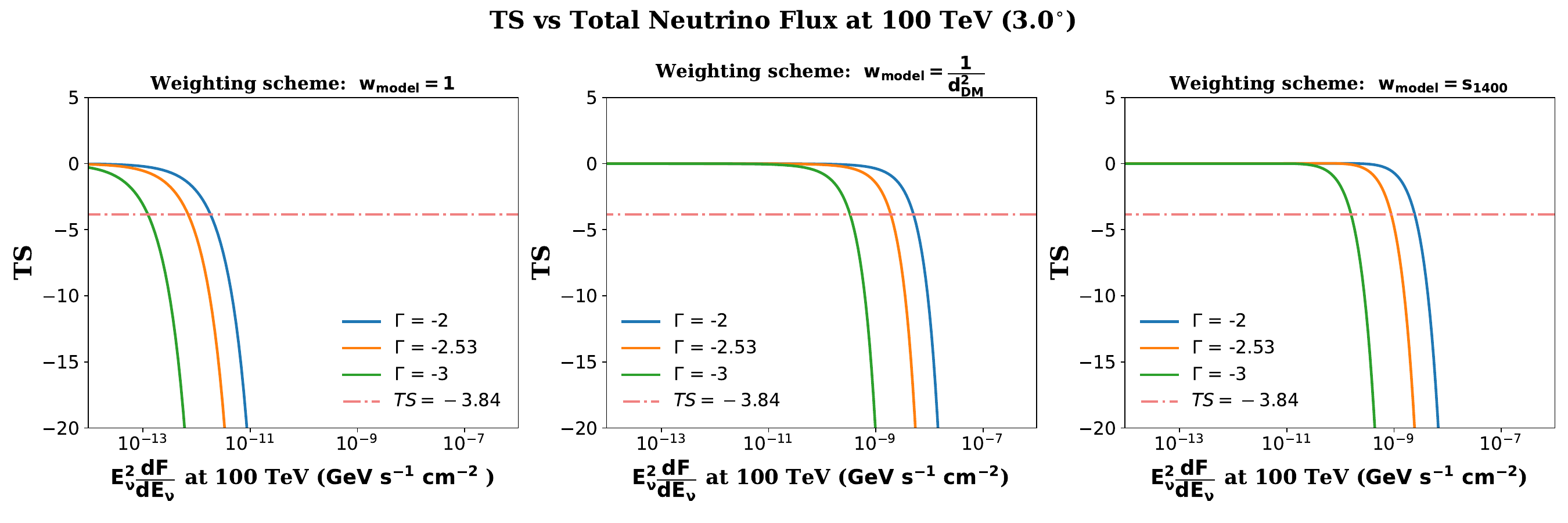}
    \caption{\label{fig:TS_C3} Plot of $TS$ (defined in Eq.~\ref{eq:TS}) as a function of the total neutrino flux for a declination band  of $\pm 3^\circ$.} 
\end{figure}

\begin{figure}[H]
    \centering
    \includegraphics[width=\textwidth]{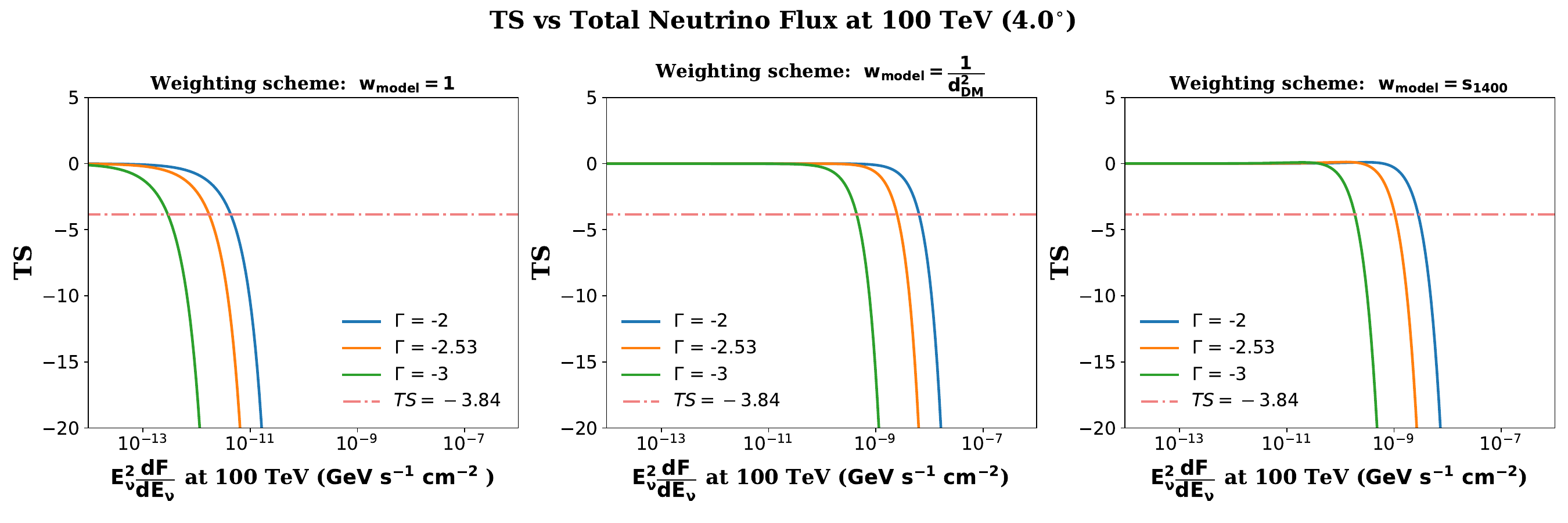}
    \caption{\label{fig:TS_C4} Plot of $TS$ (defined in Eq.~\ref{eq:TS}) as a function of the total neutrino flux for a declination band  of $\pm 4^\circ$.}
\end{figure}

\begin{figure}[H]
    \centering
    \includegraphics[width=\textwidth]{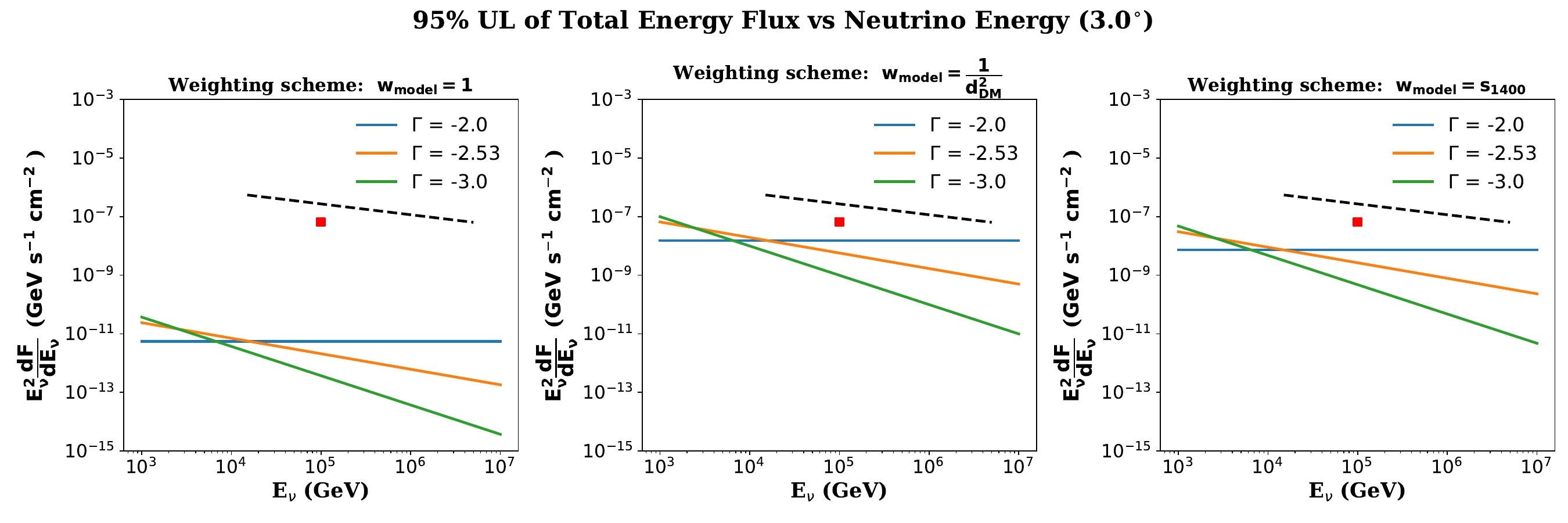}
    \caption{\label{fig:UL_C3} 95\% c.l. upper limit on differential neutrino energy flux as a function of neutrino energy for three different values of $w_{model}$ and for three different spectral indices for a  declination band of $\pm 3^\circ$.} 
\end{figure}

\begin{figure}[H]
    \centering
    \includegraphics[width=\textwidth]{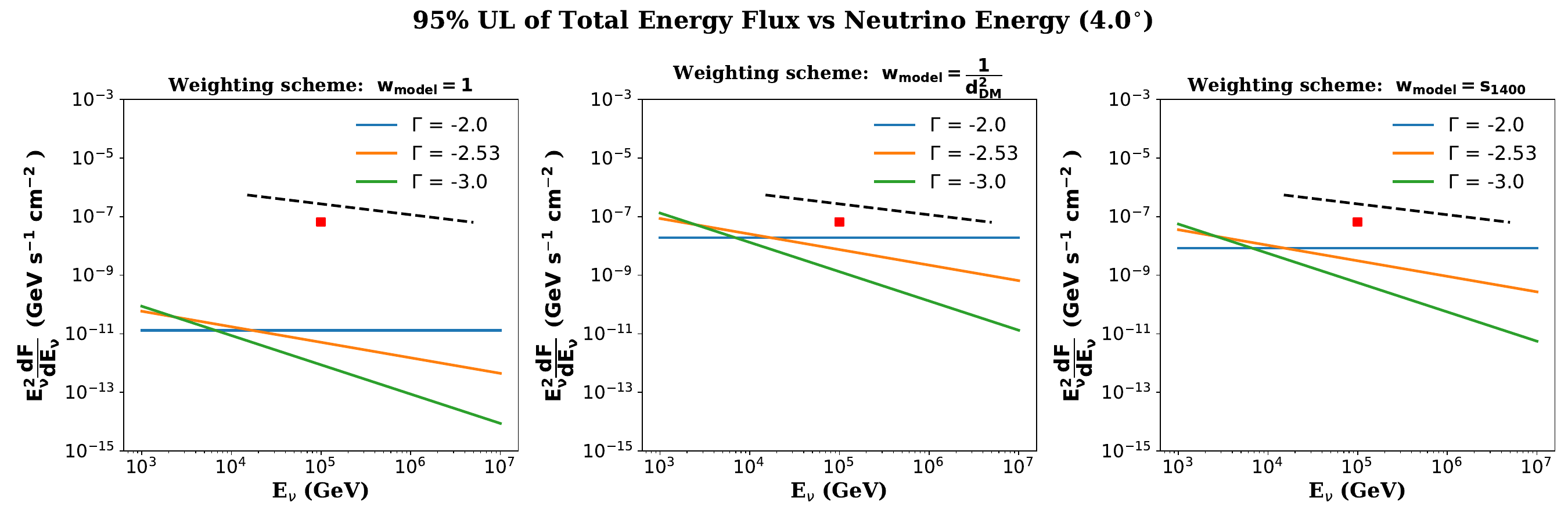}
    \caption{\label{fig:UL_C4} 95\% c.l. upper limit on differential neutrino energy flux as a function of neutrino energy for three different values of $w_{model}$ and for three different spectral indices for a declination band of $\pm 4^\circ$.} 
\end{figure}

In order to test whether our results change with different  declination 
 bands, 
we repeated the analysis in Sect.~\ref{sec:analysis} by changing the declination band  mentioned in Eq.~\ref{eq:Bi} and found no significant excess ($\sqrt{TS_{max}}>5$) in any case.  These plots can be found in Figure~\ref{fig:TS_C3} and Figure~\ref{fig:TS_C4}
for $3^{\circ}$ and $4^{\circ}$, respectively.
We then proceeded to plot the 95\% c.l. upper limit on the differential neutrino energy flux.  These plots can be found in Figure~\ref{fig:UL_C3}, and Figure~\ref{fig:UL_C4} 
for $3^{\circ}$ and  $4^{\circ}$, 
respectively.
These upper limits increase very faintly with the increase in the  width of the band and reach a maximum of less than $10^{-7} \rm{GeV~s^{-1}~cm^{-2}}$ at 100 TeV for $w_{model} \propto  S_{1400}$.

\section*{Appendix B: Tests with injected neutrinos from pulsars}
In order to provide a proof of principle test that  our analysis procedure would be sensitive to claim a  detection in the case of real signal, we validate our method by carrying out numerical experiments using injected neutrinos from pulsar positions. This is similar in spirit to the analysis carried out in Ref.~\cite{Kamionkowski}.

We show  that  the contributions from even 30 uniformly sampled sources out of 3389 sources would have been detected in our analysis. For each season of IceCube we generate synthetic neutrinos in 30 pulsars using a method similar to~\cite{Kamionkowski}. We then do the analysis with all the 3389 pulsars. For a given energy ($E$) and source declination ($\delta_j$), the number of neutrinos associated for the season $k$ and source $j$ is given by:
\begin{equation}
    N^k (E_{\nu}, \delta_j) = t_k A^k _{eff}(E_{\nu}, \delta_j) E_{\nu} \dfrac{dF}{dE_{\nu}}
\end{equation}
where $\dfrac{dF}{dE_{\nu}}$ is given in Eq.~\eqref{eq:flux}. 
We utilize $\phi_0 = 4.98 \times 10 ^{-18} \rm{~GeV^{-1} cm^{-2} sec^{-1}}$ (similar to that used in ~\cite{Kamionkowski}) and $\Gamma = -2.53$, the same parameters used to measure the total diffuse neutrino flux.

\begin{figure}
\includegraphics[width=0.4\textwidth]{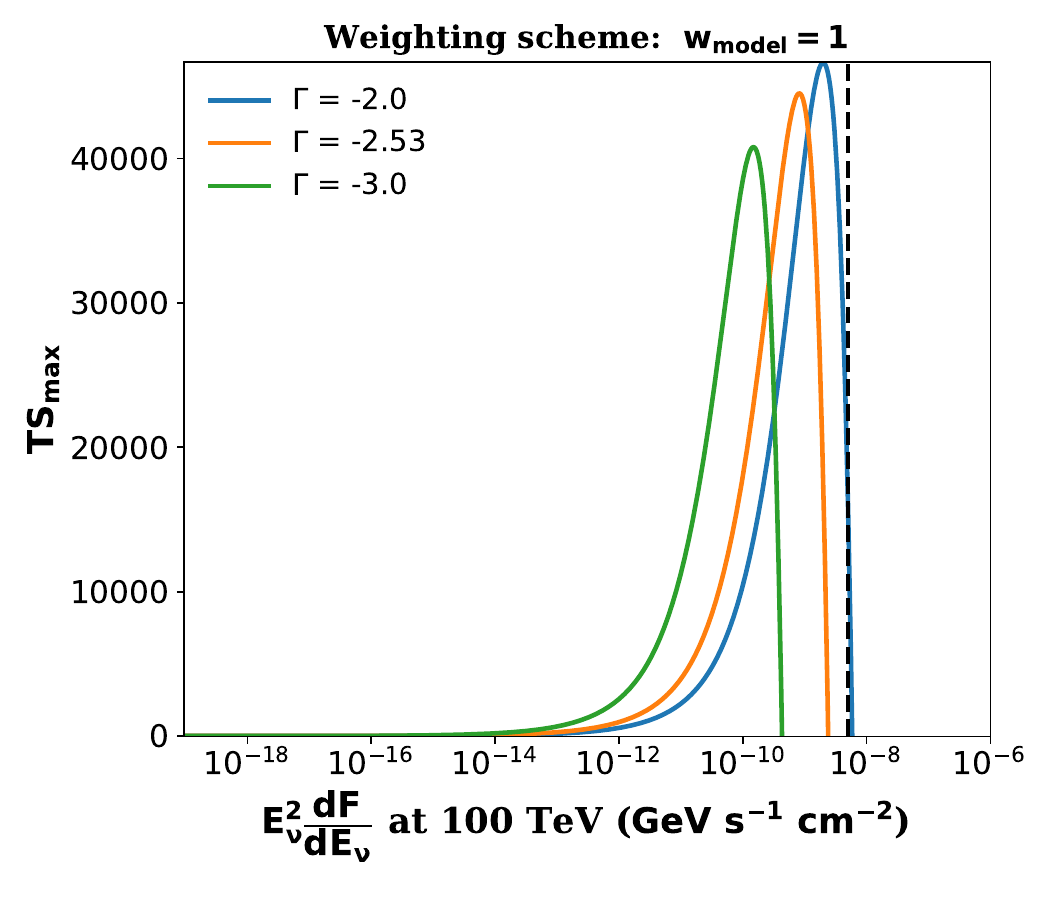}
    \caption{\label{fig:TSsignal} Plot of $TS$ as a function of the total neutrino flux for the  injected signal flux corresponding to   $\phi_0=4.98 \times 10^{-18}~\rm{GeV^{-1} cm^{-2} s^{-1}}$. The vertical  line shows  the value of the injected $\phi_0$.}
\end{figure}

\begin{figure}[H]
    \centering
    \includegraphics[width=0.5\textwidth]{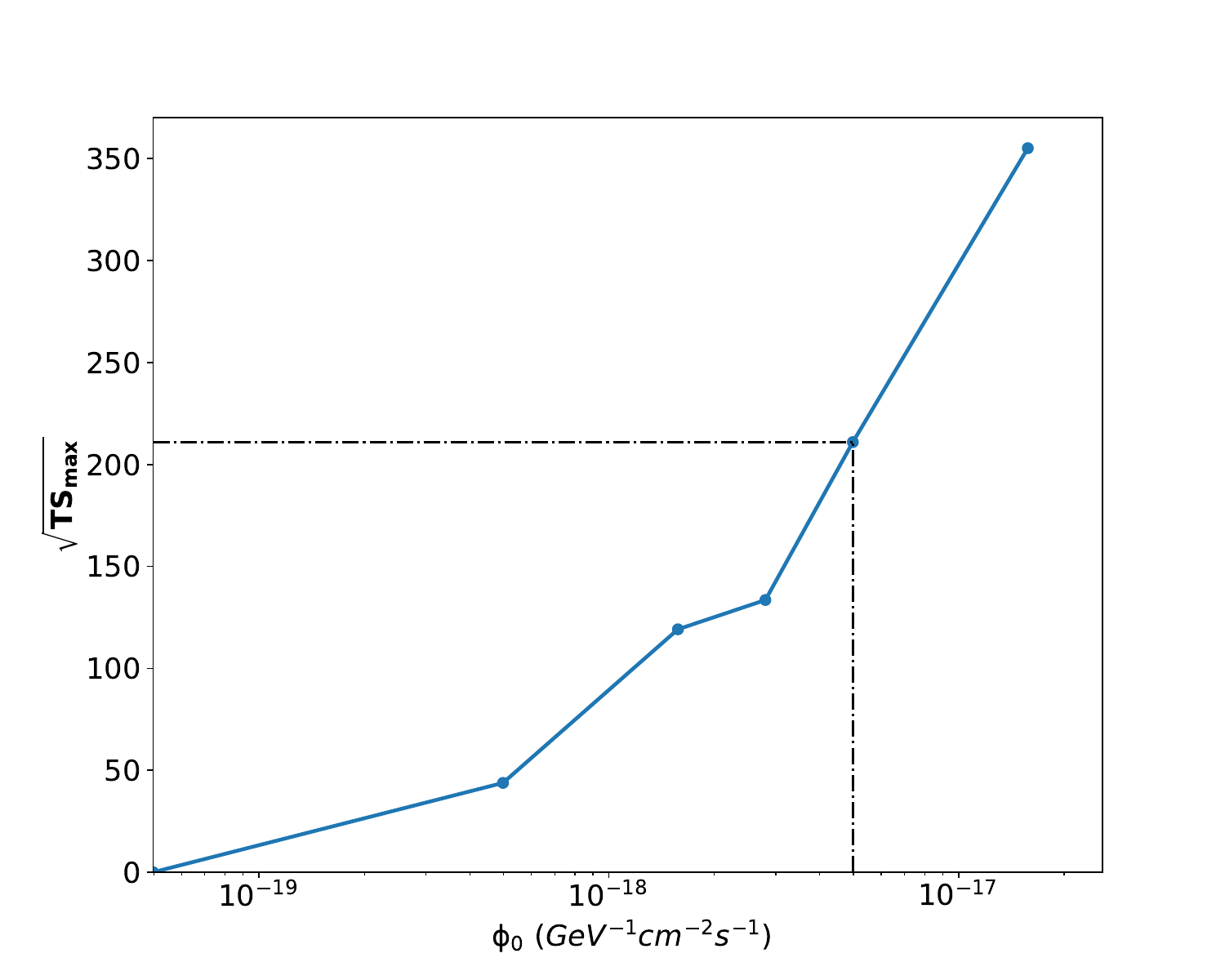}
    \caption{Significance ($\sqrt{TS_{max}}$) corresponding to the spectral index $\Gamma = -2.53$ as a function of injected neutrino flux $\phi_0$. The dashed-dotted line corresponds to the injected flux $\phi_0=4.98 \times 10^{-18}~\rm{GeV^{-1} cm^{-2} s^{-1}}$ used to reproduce Fig.~\ref{fig:TSsignal} and corresponds to $\sqrt{TS_{max}}=211$.} 
    \label{fig:TSvsphi}
\end{figure}

For each energy $E$, these associated synthetic neutrinos are smeared using an angular offset originating from a Gaussian distribution having mean equal to  0 and a variance calculated from the quadrature sum of the point-spread function (PSF) and  angular uncertainty in the reconstructed  muon direction which we assume are uncorrelated. Here, the PSF refers  to the angle between the parent neutrino  and the muon direction. Both the angular reconstruction error as well as the PSF have been made available (in the public data release) for every season of IceCube as a function of neutrino energy and declination. Some of these example  PSF plots as a function of declination have been provided in Fig.~3 of Ref.~\cite{IceCubedata}.
We inject these smeared signal events into the original IceCube data and undertake a stacking analysis with equal weighting scheme as elaborated in Section~\ref{sec:analysis} over the amalgamated data. This procedure is reiterated for different values of $\phi_0$. A plot of $TS$ as a function of the differential neutrino flux for one such value of $\phi_0$ can be seen in Figure~\ref{fig:TSsignal}. One can clearly see a statistically significant excess for $TS$  followed by a decline. We also plot the resulting significance ($\sqrt{TS_{max}}$)  as a function of  $\phi_0$ in Figure~\ref{fig:TSvsphi}. However the observed flux corresponding to the peak  in $TS_{max}$ for $\Gamma=-2.53$ is about five times smaller  than the injected flux. This is most likely due to the fact that the signal neutrino events have only been injected in a small subset of pulsars.

It is evident from  Figure~\ref{fig:TSvsphi} that the significance scales   linearly with  the logarithm of the  injected flux,
crossing the $\sqrt{TS_{max}} = 5$ threshold (corresponding to $5\sigma$ detection significance) for $\phi_0 = 2.7 \times 10^{-19} \rm{~GeV^{-1}~cm^{-2}~s^{-1}}$.
This observation provides strong evidence supporting the sensitivity of our analysis to contributions from a minimal subset of 30 sources.  
Therefore, this validates the capability of our approach to detect any putative signal from existing pulsars.

\end{document}